# An Interdisciplinary Perspective of the Built-Environment Microbiome

**Authors**:

McAlister, J.S.[1,2], M.J. Blum[3], Y. Bromberg[4,5], N.H. Fefferman[1,2,3], Q. He[6,7], E. Lofgren[8], D.L. Miller[9], C. Schreiner[3], K. Selcuk Candan[10], H. Szabo-Rogers[11], and J. M. Reed[12,*]

1. Department of Mathematics, University of Tennessee, Knoxville, TN, USA
2. National Institute for Mathematical and Biological Synthesis, University of Tennessee, Knoxville, TN, USA
3. Department of Ecology and Evolutionary Biology, University of Tennessee, Knoxville, TN 37996-3140, USA
4. Department of Biology, Emory University, Atlanta, GA, USA
5. Department of Computer Science, Emory University, Atlanta, GA, USA
6. Department of Civil and Environmental Engineering, The University of Tennessee, Knoxville, TN, USA
7. The University of Tennessee, Institute for a Secure and Sustainable Environment, Knoxville, TN, USA
8. Paul G. Allen School for Global Health, Washington State University, Pullman, WA, USA
9. One Health Initiative, University of Tennessee, Knoxville, TN, USA
10. School of Computing and Augmented Intelligence (SCAI), Arizona State University, Tempe, AZ, USA
11. Department of Anatomy, Physiology and Pharmacology College of Medicine, University of Saskatchewan, Saskatoon, SK, Canada
12. Department of Biology, Tufts University, Medford, MA, USA

* Corresponding Author: J. Michael Reed, Department of Biology, Tufts University, Medford, MA, USA, email: Michael.Reed@tufts.edu

**Abstract**

The built environment provides an excellent setting for interdisciplinary research on the dynamics of microbial communities. The system is simplified compared to many natural settings, and to some extent the entire environment can be manipulated, from architectural design, to materials use, air flow, human traffic, and capacity to disrupt microbial communities through cleaning. Here we provide an overview of the ecology of the microbiome in the built environment. We address niche space and refugia, population and community (metagenomic) dynamics, spatial ecology within a building, including the major microbial transmission mechanisms, as well as evolution. We also address the landscape ecology connecting microbiomes between physically separated buildings. At each stage we pay particular attention to the actual and potential interface between disciplines, such as ecology, epidemiology, materials science, and human social behavior. We end by identifying some opportunities for future interdisciplinary research on the microbiome of the built environment.

**Introduction**

The "built environment" comprises urban design, land use, and the transportation system, and encompasses patterns of human activity within this environment (Handy et al., 2002). The microbiome of the built environment refers to the collective community of bacteria, fungi, viruses, bacteriophages and prions, present in human-made structures, such as buildings, homes, offices, hospitals, and transportation systems. These microbiomes harbor a range of members originating from various sources – human occupants, outdoor air, water systems, soil, and even building materials. Importantly, a microbiome is more than just the sum of its individual component microorganisms. Its members interact with one another and with the surrounding environment in a cooperative, competitive, or neutral manner collectively forming a dynamic ecosystem.

Recent pandemics have highlighted the importance of where and how pathogens thrive in the built environment when hosts are present. Although the basic dynamics of some aspects of this system are well understood (Dietz et al., 2020, Pinter-Wollman et al., 2018), there is much to be gained by studying the microbiome of the built environment in an interdisciplinary setting. Those interested in the built environment microbiome from the human health perspective would benefit from interventions that could be informed by a wide range of fields, including structural engineering and HVAC systems engineering. Those who approach this topic from an environmental microbiological perspective would benefit from the building and data management perspective to understand how the environment is being used by humans.

Some work has been done to understand the microbiome of the built environment. For example, Kembel et al. (Kembel et al., 2012) found that humans have a guiding impact on the microbial biodiversity in buildings, both indirectly through the effects of architectural or engineering design, and more directly through the effects of human occupancy and use patterns in different spaces and space types. A key finding of this work – the fact that source of ventilation air has the largest impact on bacterial diversity – has been confirmed by other studies (Meadow et al., 2014). These results suggest that we can alter indoor microbiomes, selecting the microbial species that potentially colonize humans during our time indoors.

Even though many different bodies of literature have explored individual facets of the microbiome in the built environment, it is not always obvious how their findings can be integrated. Being able to bridge these gaps will improve the study of the built-environment microbiome in every discipline. With a better synthesis of the field we will be able to understand and evaluate risks as ecological processes. We will be able to design better powered, more informative, and more targeted studies to understand the multifaceted nature of the microbial built environment. Finally, we will be able to optimize mitigation strategies based on a more complete and holistic understanding. Improving the interconnectedness of the field improves the ability of every researcher in every discipline to refine and advance their work.

We are not attempting to provide all the necessary tools of collaboration in this overview. For a comprehensive discussion consider the National Academies report (2017). We instead present an overview of the elements contributing to the ecology of how microbes function within the built environment in order to synthesize ideas about how we understand the microbiome, how we measure it, and how it changes in time. We are not focused specifically on human health, but as much of the work on the microbiome in the built environment comes from this field, we rely on it for our general link to microbial ecology (Mohajeri et al., 2018).

Here we highlight examples of how the built environment can affect the basic ecology and dynamics of microbial communities. Rather than the traditional human focus of microbes in the built environment, we approach the microbiome-built environment interface from the microbial perspective. We focus primarily on a within-building microbial ecology framework, but we end by presenting a landscape-scale (between buildings) perspective. This overview and synthesis of built environment microbiomes will allow for the creation of a modeling framework that can help to describe, and ultimately predict, the microbiomes of particular built environments.

**The Microbiome Ecology – Built Environment Interface**

Human-designed and -built environments are meaningfully different in many ways from natural environments. These differences have the potential to foster the growth of profoundly different microbes and the establishment and organization of profoundly different microbial communities. As with natural environments, the physical structure and system processes (functions) of the built environment affect the ecology and dynamics of microbial communities. These communities are further affected, both directly and incidentally, by design features specific to the function of the built environment (housing, hospital, etc.), as well as by human activities. In fact, in both natural and built environments, habitat manipulation provides options for species management. In the built environment, architectural design and engineering can directly affect the microbial communities present and which types of activities are likely to be undertaken, including those to reduce risks to human health (e.g., D'Accolti et al., 2022; Gottel et al., 2024). In addition, there can be incidental impacts on the microbiome when architectural design focuses on goals beyond simple function, such as increased energy efficiency or facilitating human interactions (social or work-related) (Shrubsole et al., 2014; Heida et al., 2022).

**Niche space and population refugia**

What a species does, as well as where and how it does it, defines its niche in an ecosystem ( Kembel et al., 2012; Carscadden et al., 2020). The microbiome of the built environment is depauperate compared to that in natural communities, in part because the built environment is structurally less complex. Regardless, ecological studies of the natural environment provide a natural parallel for investigating the built environment. For example, manufactured structures create potential niche space for species – in our case microbes – that differ in many of the same characteristics as in the natural environment, including differences in physical space, isolation, light, humidity, moisture level, temperature, accessibility, etc. (Hao et al., 2024). These features affect the types of species that can colonize and establish in each site (space within a building), and consequently through interspecific interactions, the community composition and structure (Kembel et al., 2014).

Refugia for microbes occur in a variety of predictable places in the built environment, i.e. those associated with waste disposal, standing water (or moisture in general), air-transport systems, and in sites that are relatively inaccessible to cleaners or are not well maintained (Nazarenko et al., 2023). For example, Legionella outbreaks can occur when water from poorly maintained cooling systems create a refuge for bacterial growth; in this case, rather than an HVAC subsystem filtering the pathogen, it acts as a centralized source of contamination (e.g., Prussin et al., 2017).

Engineering designs and building functions affect the amount and type of niche space available and can be altered to minimize microbial opportunities. For example, in hospitals and veterinary clinics efforts are made to eliminate the accumulation and spread of microbial pathogens (Wright et al., 2008; Assadin et al., 2021). Both human and veterinary healthcare settings have particular context-specific concerns surrounding the microbial built environment, primarily centered on the pathogenic microbial communities. By their very nature, healthcare environments are full of patients who are likely to be shedding pathogenic microbes into the environment, providing a ready source of new importation. For example, a significant amount of engineering work has gone into designing air circulation and filtering systems that minimize the spread of pathogens, particularly in hospitals (Beggs, 2003; Bolashikov and Melikov, 2009), although lessons have expanded to other built environments (e.g., Arjimandi et al., 2022). Similarly, the design of daycare facilities may incorporate accommodations for distinct types of interactions between human occupants and designed structures; here, normal anticipated use involves more mouthing and chewing of communally accessible surfaces as well as more contact with floors than would normally be considered advisable (Reed et al., 1999).

Despite considerable gains, lingering questions about how to effectively "harden" the acute care environment against microbial contamination as well as how to control pathogens within that environment remain. Functionally, removing microbial habitat (whether by engineering design or through effective cleaning; e.g., Edwards et al., 2019) and/or altering viable routes for dispersal alters microbial diversity, abundance, and persistence (Walters et al., 2022).

**Mapping the Microbial Environment**

A crucial part of understanding the microbiome of the built environment is understanding how it is measured. Observing the microbiome and building a map of the microbial environment is important from a public health perspective (Kim et al., 2020 ;Shi et al., 2021). It allows for real time assessment of risk to humans and, with multiple data points, it can inform decisions about design and utility of the built environment. Moreover, mapping the microbial environment is a crucial first step for using predictive modeling (e.g., Pasarkar & Pe'er, 2021). Without an understanding of what is in the microbiome and where it is distributed throughout the built environment, even the most accurate and sophisticated predictive models will fail to have predictive power. For this reason we present here two ways to think about observing the microbiome of the built environment

The first approach is marker-based tracking (e.g., Tedersoo & Lindahl, 2016). It is common to use various markers - either inert chemicals that can be detected, such as gels that glow under UV light or benign microbes – to map the microbial environment, especially but not exclusively in healthcare settings. At the most basic level, this is done to ensure that cleaning and disinfection procedures are successfully being followed - marker compounds or organisms should be removed if procedures are being followed correctly (Miranda et al., 2011). More generally however, this can also be used to establish pathogen movement. For example, sampling human-touch surfaces in a veterinary hospital for Methicillin-resistant *S. pseudintermedius*, a pathogen in companion animals that rarely infects humans, was used to indicate contamination of multiple surfaces within veterinary hospitals (Feßler et al., 2018). Surrogate markers for microbial contamination, such as cauliflower mosaic virus, have been used extensively to demonstrate the potential movement of microbes within healthcare environments, from stethoscopes and clothing to portable equipment in hospitals (e.g. Jiang et al., 1998).

      The second mapping approach uses metagenomic understanding of the microbiome. Until recently, microbiome analysis most frequently referred to the exploration of the microbiome member bacterial species, as in the marker-based tracking mode. The process for identifying "who" was present in a particular microbiome included amplification and sequencing of the various variable regions of the 16S rRNA gene (RNA of the 30S ribosome subunit) – the gene proposed by Woese et al. (Woese, 1987; Woese et al., 1990) as a molecular marker of prokaryotic evolution. While exceedingly useful in describing evolutionary processes, 16S rRNA sequences are limited in precisely identifying the organisms they come from. With the sharp drop in costs of sequencing, metagenomics, i.e. whole metagenome sequencing (see below), has become much more common. While metagenome analysis can answer the question "who is there?" it also allows a more complete look at "what are they doing?" By establishing the molecular functionality encoded in the metagenome directly using analyses of DNA data obtained via high-throughput sequencing technologies (Mora et al., 2016), it is possible to bypass the assumptions that microbiome members are essentially the same as individual culturable microbes, as well as forgo the error-prone process of genome assembly and organism mapping biased by incompleteness of databases.

      Mapping the built environment poses qualitatively similar challenges to mapping the natural environment, such as tracking down the often-hidden reservoirs of microbes (Adams et al., 2015; Christoff et al., 2019). This requires regular monitoring of the entire built environment and engineering designs that allow accessibility to potential problem spots, i.e., new or repeating microbe reservoirs on invasion conduits. Although swabbing sites is the most common collection method, others are being developed, such as using condensation traps (Hampton-Marcell et al., 2023).

**Population dynamics**

From the perspective of a population ecologist, the microbiome, like any biome, can be thought of as the collection of coexisting microbes in a particular physical space, where a population ecologist would be interested in the dynamics of one or more of the taxa. For the target microbe, their distribution in a built environment is probably not continuous; rather, it will be patchily distributed. The amount of movement between patches determines whether all the individuals constitute a single population (extensive movement), multiple populations (isolated), or a metapopulation (numbers driven by local dynamics, with local extirpations and recolonizations) (Smith and Green 2005; Fink and Manhart 2023) provide a perspective of the dynamics of microbial populations in natural settings. Some of highlights that make microbial population dynamics fundamentally different from that of, say, terrestrial vertebrates, is their capacity for rapid population growth, with doubling times measured in hours or days, and the small absolute spatial scale of their growth patterns but comparatively large scale across which they can disperse.

Although qualitatively the concepts of traditional population biology are also applicable to microbial populations, there are limitations. Two difficulties identified by Fink and Manhart in investigating microbial population dynamics are the difficulties in determining absolute abundances (researchers are currently restricted to relative abundances) and the difficulty understanding short-term dynamics because of insufficient sampling frequency. An alternative to time series investigation of populations that has been proposed is determining instantaneous growth rates, but this has not had much success in natural populations (Carroll et al. 2022). So, application of population models to microbial populations is still limited relative to that of vertebrate population dynamics.

How will the dynamics of microbial populations in the built environment differ from that of natural populations? One might imagine that the relatively simpler communities in the built environment might make understanding their dynamics simpler, converting to a relatively smaller set of primarily human-dominated microbes following construction (Gaüzère et al. 2014), but human interventions (like cleaning) can make the populations less stable (Young et al. 2023).

Built environments have predictable compartmental structure, atmospheric controls, occupation patterns, specific utility, high immigration and emigration, as in transportation hubs, intense selective pressures depending on the function of the built environment, and artificial mechanisms of dispersal, as in plumbing or HVAC systems (Gilbert and Stephens 2018). As we gain understanding of the ecological requirements of microbial species (Krueger, 2019), and how they interact with the particular features of a built environment and human interventions, we anticipate improved predictive capacity for microbial population dynamics.

As an example, a built environment such as a hospital can be thought of as a metapopulation of a room-level community within an ecosystem, with movement between communities being equivalent to human movement between rooms via corridors. This conception allows population ecologists to make predictions about microbial communities in the built environment and to illustrate the importance of hand hygiene and PPE (Lofgren et al. 2016). Another such example is the analogy between *C. difficile* and fluoroquinolone antibiotics and invasion ecology after a catastrophic event (Waaij 1989), where ecological interactions are perturbed and the progression through the transient states after the perturbation can lead to eventual arrival at a different equilibrium. Combined with the existing understanding of invasion and succession, we anticipate advancing our understanding of microbial population dynamics of the built environment through population modeling, with expectations similar to those realized by modeling disease systems (e.g., Kopec et al. 2010, Tatem et al. 2012).

**Metagenomics – microbial community ecology**

Another distinct but equally valuable approach to understanding the microbiome of the built environment is through the study of community ecology, which is captured using metagenomics. That is, identifying the microbiome structure (taxa/species, relative abundances) and function (ecological) using DNA sequencing of samples from the environment (Wooley et al. 2010). The metagenome comprises a vast array of genetic material that encodes functional genes and pathways (Singh 2009, New and Brito 2020) and the built environment shapes the composition and characteristics of its microbial inhabitants.

While each microbe brings to an environment its own genetic material and metabolic capabilities, member interactions guide total metabolic capacity. Furthermore, synergistic relationships may emerge, where the presence of certain microbes enhances the survival or growth of others, thereby changing genetic content as well. One of the best studied examples of such synergies is that of keystone species that, incidental to their local dynamics, alter environmental conditions to facilitate colonization by others. For example, cross-feeding, i.e., the exchange of vitamins, amino acids, and nucleotides, is common across bacteria (D'Souza et al. 2018). However, keystone species may also alter other factors, such as metabolic regulation (Tudela et al. 2021). Bacterial interactions also suggest emergent functionality, i.e. molecular functions, available to the community, but not individual microbe (Chung et al. 2024).

What might we expect of the metagenome of the built environment? As mentioned above, the microbiome of the built environment is simplified compared to that of natural microbial communities, yet more dynamic because of human actions and interventions. One possible result of these occurrences is that population and community dynamics might be transient, rather than existing in stable states (Fujita et al. 2023). Consequently, the microbial community might be more difficult to characterize (because it has limited stability) and surface sampling to investigate the microbiome (e.g., Perkins et al. 2022) might need to be more frequent than otherwise expected to track changes over time. This also might reduce the predictability of community responses to building alterations, changes in human activity, or interventions.

A clinical conception of the built environment also allows for designing spaces to effectively monitor pathogens – for example, the placement of plumbing in such a way as to allow potentially targeted wastewater monitoring as well as to mitigate spread, for example, by allowing for spacing and distancing needs to be considered in the design phase, improving ventilation, or providing opportunities for hand hygiene in areas where pathogen burdens are likely to be strongest (Dai et al. 2017, NAS 2017). While narrow in its ecological scope, the clinical conception of the built environment microbiome allows clinicians and researchers to optimize built environments for safety and functionality.

From a practical point of view, the physical distribution of microbes in the built environment, as well as expectations of community structure and function, are driven by numerous on-site factors. For example, if the space is used differently than intended, such as turning a bank into a fast-food restaurant, the high-touch or dirty areas are likely to be very different (e.g., a food waste site where none had existed). Human behavior can also alter the microbiome: space designed for one purpose may be used contrary to its original design, such as overcrowding or temporarily using a school gym as a make-shift hospital (Turroni et al. 2017). Finally, there is a plethora of problems the built environment can experience that alter the microbiome, such as architectural design failures (e.g., inadequate drainage), function failures (e.g., power outages, water supply disruption, HVAC failure, and disasters (e.g., Smith and Casadevell 2022)), as well as building degradation (e.g., concrete breakdown), all of which have the potential to alter substrates, colonization potential, and microhabitats (Kiladel et al. 2021).

**Dispersal and colonization**

The built environment microbiome rapidly transitions to reflect its human inhabitants (Young et al. 2023). Ignoring, for now, movement between built environments, there are many ways in which microbes can move within the built environment, and the different methods of transmission can affect population dynamics and metagenomics. These have been of interest to clinicians interested in human health, and their knowledge will help us understand the built microbiome more broadly. While in large, open areas, such as atria or enclosed arenas, a diffusion model might be sufficient (e.g., Scott et al. 1995), in a built environment there may be mechanisms that affect microbe dispersal that require specific consideration. Here we briefly review broad microbial transmission mechanisms - dispersal and colonization - within the built environment. Each could be modeled separately to predict microbiome dynamics in a particular built environment.

*Aerial Dispersal* – Microbes can be transported through the air by a variety of mechanisms. Air systems, such as HVAC, are fundamental drivers of circulation and exposure (Burge 2018, Sodiq et al. 2021). Unique to the built environment is the impact of HVAC systems on the way pathogens and other microbes are transported through the air. HVAC systems impact the microbiome by altering the temperature and humidity of the circulating air (Walther and Ewald 2004, Lin and Marr 2019) and they determine how long pathogens remain suspended as aerosoles or droplets before settling/falling onto surfaces (Drossinos and Stilianakis 2020). The existence of these systems represents a control on the microbiome which is uncommon outside of the built environment. Sub-HVAC systems, such as filters or purifiers, are meant to extract contaminants, including pathogens from the circulating air (Nazarenko et al. 2023), but can also themselves act as centralized sources of contamination (e.g., Purssin et al. 2017). We also note that dispersal is affected by the degree to which a building is sectored, such as having HVAC systems that separate, for example, human and animal ventilation systems.

Exhaled air, expelled directly from infected hosts, also drives microbial dispersal (de Oliveira et al. 2021; Walker et al. 2021). Combined with corresponding inhalation creates a net effect of a complicated source-sink dynamic (Roy et al. 2010). Ultimately, the fate of the inhaled pathogens is dependent, in part, on the effectiveness of the innate and adaptive immunity of the host, as well as the tissue tropism of the pathogen and/or its community (Kim D. et al PNAS 2020) .

Of course, these sets of factors are not independent of each other – HVAC systems alter the spatial dynamics (and therefore patterns of exposure) of exhaled air (Zhang et al. 2019). Sub-HVAC systems are specifically designed to work between the HVAC and respiratory layers, but also directly impact HVAC performance (Feng 2019) and concomitant pathogen transmission risks (Duill 2021) throughout a building. HVAC-driven alterations in air can even impact the physiological processes of exhalation (Yang and Marr 2011) and susceptibility to exposure (Mäkinen et al. 2009). Additionally, the complicated spatial structure of the built environment creates a highly interconnected network or patches, each with their own parameters for uptake into the air, filtration out of the air, deposition onto surfaces and into water.

*Water Dispersal* – The distribution of water in the built environment is highly engineered to minimize contamination of potable water and to effectively remove waste water from the built environment. However, water can still provide a way for microbes to be transmitted throughout a building and provides a crucial reservoir for some parts of the microbiome.

Premise plumbing (transport system for water throughout a building) is characterized by elevated temperature, diminished disinfectant concentration, prolonged stagnation, and increased biofilm growth, making it an ideal ecological niche for opportunistic establishment, growth and dispersal of pathogens, such as *Legionella*, *Mycobacterium*, and *Pseudomonas*. As a result, bacterial levels in premise plumbing systems can be orders of magnitude higher than in the water main (Li et al. 2021). Often protected by biofilms, these communities can readily disseminate throughout a building and are often extremely difficult to control due to the protective nature of the biofilm itself (Maillard and Centelenghe 2023). In addition, contaminated moisture leaking into a built environment provides a pathway for microbes to be aerosolized and become transmitted aerially .

*Human Occupancy Dispersal* – Humans affect microbial dispersal in a built environment by affecting the spatial distribution of microhabitat and by actively transporting microbes. Human presence at different densities alter temperature and humidity, which change habitat suitability profiles (Qiu et al 2022). Physical contact involved in human use of the space (e.g., sitting on chairs, leaning against walls, etc.) can disrupt spatial patterns in microbial colony growth and also introduce novel microbes into an otherwise established system (Lopez et al. 2013, Stephens et al. 2019, Wang et al 2022). Concomitantly, contact can reduce existing populations of microbes by transference from the environment to the humans who then carry them (either passively or under active ongoing replication) to other locations (Zhang et al. 2021). Humans also actively clean areas of their environment, though frequently in response to visible stimuli (e.g., dirt) rather than in direct response to microbial activity (Campkin and Cox 2012). Even large numbers of people moving down relatively narrow corridors can transport microbes in their wake (Jha et a. 2021).

We also note that within a building there is human-mediated dispersal of microbes through 'hitchhiking' on people, food, or goods, and that these movement patterns can be centralized or decentralized. Food services, for example, tend to be centralized, with a single source of food either radiating outward, or people moving centrally to get food. In contrast, the movement of some goods, like wheelchairs or CPAPs to wherever they are needed is decentralized. These different dispersal patterns will differentially affect recolonization of cleaned surfaces, and of standing microbial communities.

Of course, each of these examples relates to the extrinsic interface between humans and their environment. Humans also harbor diverse and complicated microbial communities within their bodies and have multiple pathways for shedding species into the environment, facilitating microbe dispersal (Stein 2011). While much work has been done to characterize rates of bacterial shedding for a variety of pathogens in veterinary medicine (Roberts and Crisler 2005; Subharat 2010; Chen et al. 2013; Krebs et al. 2023), very little work has been done studying rates of replication and shedding for non-pathogenic bacteria, and even less has been done when restricted to those carried on/in humans.

One of the main purposes of architectural design is to guide humans through spaces in a manner that encourages appropriate and efficient use of the space provided. Narrow hallways that can become bottle-necks to traffic flow are less likely to contain benches than wider atria, meant to encourage gathering and leisure. These use cases also affect how humans impact the microbial communities of each region of the built environment. Areas built to encourage lingering of large groups (e.g., atria, open floor plan cubicle offices, etc.) will likely encourage a different microbial community from those that foster maintained presence from a more limited number of humans (e.g., private offices, small meeting rooms, etc.), which again will likely differ meaningfully from shorter duration use, but high throughput areas (e.g., elevators, office kitchens, restrooms, etc.). While the patterns of flow have been well studied, their implications for how those use patterns result in distinct microbial communities is less well explored. The majority of such studies have occurred in the context of infection control in healthcare settings (Anderson et al 2018; Rutala et al. 2018; Hajime et al. 2021).

Beyond human-mediated dispersal, there are also a variety of human-adjacent animal mediators of similar phenomena. Companion animals and urban pests such as rats, mice, or cockroaches are also likely to affect microbial communities in similar ways, albeit via different precise routes through the built environment. Engineered design occasionally does consider how best to discourage pests, but to the best of our knowledge, does not consider the additional complexity of accounting for the impact of their presence and movement on the microbial community of the environment.

**Evolution**

The evolutionary capacities and mechanisms of microbes have been reviewed before (e.g., Kessell 2013; Morschhäuser et al. 2000; Brennan and Logares 2023), so here we will limit our comments to ways in which microbial evolution might be modified by the built environment.

The built environment is selective, shaping the composition and characteristics of its microbial inhabitants. That is, over time, microorganisms within the built environment can adapt and evolve to better thrive in these human-made habitats. Certain microbes may develop specialized traits or mechanisms to withstand environmental stresses, resist antimicrobial agents, or use novel resources. This evolutionary process contributes to the ongoing dynamics and resilience of the metagenome, i.e. the totality of the genetic information present in the microbiome.

We see two broad ways in which microbial evolution could be modified by the built environment. First, the frequency and severity with which surface cleaning is done creates strong selective pressures on the microbiome (Artasensi et al. 2021). The built environment, particularly residences, offices, event centers, etc., are cleaned regularly. In the clinically-focused literature, there are excellent studies that have considered the impact of different patterns and types of cleaning efforts (Mitchell et al. 2019), and how it might be best to tailor such efforts to the type of built environment targeted for microbial reduction (Carling and Huang 2013). Cleaning to remove microbes is a harsh disturbance that is a strong selective pressure, favoring cleaning agent-resistant microbes, such as spore-forming bacteria or those that form biofilms. In addition, if cleaning is frequent, the continued disturbance creates a selective pressure for rapid population growth and it creates an invadable surface for colonizing microbes (McDonnell 2020). If microbes show life-history characteristics parallel to larger organisms, this type of disturbance pattern would favor r-selected species - that is, those with good dispersal capacity, high reproductive rates, and short life-spans (Stearns 1976; Reznick et al. 2002).

Second, the regular clearing and reinvasion of cleaned surfaces, combined with the high opportunity for colonization associated with human intrusion rates, will create novel communities (microbiomes) and favor a high rate of mutation. This, in turn, will likely introduce novel strains and increase the likelihood of microbes with novel functionalities favored by these dynamic environments, most often acquired via lateral gene transfer (Woolhouse et al. 2005; Mohsin et al. 2021).

**Landscape ecology**

In viewing the microbiome of the built environment from an ecological perspective, we note a tremendous opportunity for drawing on the concepts and tools of landscape ecology and biogeography. It has been proposed that there is a landscape ecology of microbes in the built environment (e.g., Mony et al. 2020) although it has only rarely (to our knowledge) been formalized in any way (e.g., Pattni et al. 2023). Landscape ecology concepts have already been invoked to study microbiomes within an individual (e.g., Proctor and Relman 2017; Couch and Epps 2022); we believe that with little effort they could be scaled up spatially to the built environment. The built environment can easily be viewed as parallel to a natural landscape ecology: there are habitat patches (buildings), connected by corridors (transport systems), embedded within a matrix of non-habitat (Francis et al. 2022). The degree of connectivity between structures in a built environment includes both transportation systems, which are part of the built environment, and the degree to which people move between structures on a daily basis outside built structures.

This type of connectivity of the built environment can be modeled using a network approach (e.g., Krüger 1979), and could be applied to microbial communities. While we think this is one useful approach, there is a panoply of concepts and research tools from traditional landscape ecology that could be applied to the built-environment microbiome. Further, it lends itself well to rapid advances through modeling, from ordinary differential equations to Markov chains to spatially explicit, agent-based models.

In addition to deliberate and incidental transport of microbes between built structures, there is also the possibility of incidental - system adjacent - microbial spillover to (or from) the built environment. For instance, when considering the placement of new structures where there can be a risk of microbial (pathogen) spillover. An example is the Pirbright Institute in England which incited a foot and mouth outbreak on an adjacent farm in 2007 (Cottam et al. 2008). Taking a pathogen-specific perspective to understanding the microbial community as a whole could help inform decisions about placement and design of the built environment ranging from the landscape-level, to what materials to build and furnish a space with, and what compounds might be used to help clean it.

This highlights a tremendous opportunity to increase collaborations in built-environment projects among civil engineers, material scientists, architects, microbial and macrobial ecologists, health-care workers, and the intended end-users of new construction.

**Additional opportunities for multidisciplinary work on the microbiome of the built environment**

The built environment provides excellent opportunities to study microbial ecology via adoption of a landscape-ecological perspective to large-scale assessments of the microbiome just discussed, including integrating research across disciplines. While the opportunities are diverse and limitless, within the scope of this brief review, we outline three examples that highlight the inherently interdisciplinary scope of research in this area.

The materials used in constructing the built environment influence microbial communities and provide opportunities for pathogen reduction. For example, building materials interact with humidity and moisture to facilitate microbial establishment and growth, which differentially affects their deterioration (Gaylarde and Morton 1999). Construction materials also differ in their susceptibility to support microbial reservoirs (Munir et al. 2020; Course et al. 2021). Interfacing with materials science and engineering, an active area of research is making building materials more resistant to microbes, including creating antimicrobial concrete, nature-based antimicrobial surface structures, and surface treatments via polymers, nanotechnology, and doping with metallic ions (Qiu et al. 2020; Soni and Brightwell 2022; Kirthika et al. 2023). In an interesting twist, there is research showing that microbes might be used to decrease materials degradation (Junier and Joseph 2017), so there is much to explore at this interface.

It turns out that plants do more than just improve the psychological health of occupants of a building (Bringslimark et al. 2009) - they also affect the microbiome (Mahnert et al. 2015). At a basic level, plants provide novel microhabitats for microbes, particularly due to the presence of soil. Plants in sealed buildings increase oxygen locally, and they and their associated root microorganisms (rhizobiome) - a microbiome in its own right - can remove volatile organic compounds and some pollutants, such as ammonia and asbestos ( Aydogan and Cerone 2021). In fact, plants have been investigated as biofiltration systems to supplement air filter systems (Darlington and Dixon 2000). All of these actions and activities affect the microbiome of a built environment.

As a final example, even the social and physiological interactions with the built environment can have surprising latent interactions with the microbiome. Many buildings rely on temporally distinct shifts of people with equally distinct roles (and therefore interactions with the environment) (Mangkuto et al. 2014). For example, professional office workers may occupy spaces during the day that are occupied at night by janitorial staff, while conversely hotel rooms are commonly cleaned during the day by a regular staff member of the hotel, while the occupancy of those same rooms during the evening involves continuous turn-over. Since cleaning and janitorial activities constitute regular perturbations of microbial communities, these alternating patterns in when and by whom they are re-seeded with new microbes may have profoundly different outcomes relative to environments without this planned, regularly alternating pattern of (re)introduction. This may be further complicated by the circadian disruption endured by night work that can depress immune function (Rivera et al. 2020) and in other ways alter individual microbiomes, thus potentially shifting the distribution of likely microbes carried by the nighttime occupiers of the environment (Mortaş et al. 2020; Neroni et al. 2021).

**Conclusions**

The built environment is driven by human population density, needs, material availability, and a wide range of circumstances from careful planning to *ad hoc* construction to emergency responses. As we have tried to highlight here, there is already science addressing microbial communities in other contexts, including colonization capacity, rapid population growth, and rapid, flexible evolution, and the built environment is qualitatively similar, (2) microbiome disruption is consistent and can be modeled, and (3) all of it is important to human health. The suite of characteristics and situations found in the built environment provides ample opportunities for disasters such as pathogen outbreaks. It also provides equally ample opportunities for effective cross-disciplinary research, and resolution. Experience in many different areas of the human-natural interface has shown that multidisciplinary teams have the potential to be effective at understanding and resolving complex issues where siloed research might fail or be slower to a solution (e.g., Doyle 2008, Cuevas et al. 2012, Islam and Susskind 2012, Mooney et al. 2013, Piorkoski et al. 2021). We propose that understanding and manipulating the microbiomes of the built environment offers a suite of issues and opportunities and we hope these perspectives will help excite others to join us in pursuing them.

**Acknowledgements**: This material is based upon work supported by the National Science Foundation under grant CCF #2200140.